\documentclass[10pt,letterpaper]{article}
\usepackage[top=0.85in,right=0.5in,left=0.9in,footskip=0.75in]{geometry}

\usepackage[utf8]{inputenc}

\usepackage{cite}

\usepackage{nameref,hyperref}

\usepackage[english]{babel}

\usepackage{microtype}
\DisableLigatures[f]{encoding = *, family = * }

\raggedright
\usepackage{changepage}

\usepackage[aboveskip=1pt,labelfont=bf,labelsep=period,singlelinecheck=off]{caption}

\makeatletter
\renewcommand{\@biblabel}[1]{\quad#1.}
\makeatother

\usepackage{lastpage,fancyhdr,graphicx}
\usepackage{epstopdf}
\fancyheadoffset[L]{2.25in}
\fancyfootoffset[L]{2.25in}

\usepackage{color}

\definecolor{Gray}{gray}{.25}

\usepackage{graphicx}

\usepackage{sidecap}

\usepackage{wrapfig}
\usepackage[pscoord]{eso-pic}
\usepackage[fulladjust]{marginnote}
\reversemarginpar

\usepackage{tikz}

\usepackage{subfig}
\usetikzlibrary{shapes,arrows}

\tikzstyle{vertex}=[circle,line width=0.1mm,draw=black!80,minimum size=10pt,inner sep=0pt]
\tikzstyle{selected vertex} = [vertex, fill=red!24]
\tikzstyle{edge} = [draw,thick,-]
\tikzstyle{weight} = [font=\small]
\tikzstyle{selected edge} = [draw,line width=5pt,-,red!50]
\tikzstyle{ignored edge} = [draw,line width=5pt,-,black!20]

\begin{document}

\vspace*{0.35in}

\begin{flushleft}
{\Large
\textbf\newline{Efficient and Accurate Robustness Estimation for Large Complex Networks}
}
\newline
\\
Sebastian Wandelt\textsuperscript{1},
Xiaoqian Sun\textsuperscript{1,*}
\\
\bigskip
\bf{1} National Key Laboratory of CNS/ATM, Beihang University, Beijing, China
\\
\bigskip
* sunxq@buaa.edu.cn

\end{flushleft}

\section*{Abstract}
Robustness estimation is critical for the design and maintenance of resilient networks, one of the global challenges of the 21st century. Existing studies exploit network metrics to generate attack strategies, which simulate intentional attacks in a network, and compute a metric-induced robustness estimation. 
While some metrics are easy to compute, e.g. degree centrality, other, more accurate, metrics require considerable computation efforts, e.g. betweennes centrality.
We propose a new algorithm for estimating the robustness of a network in sub-quadratic time, i.e., significantly faster than betweenness centrality. 
Experiments on real-world networks and random networks show that our algorithm estimates the robustness of networks close to or even better than betweenness centrality, while being orders of magnitudes faster. Our work contributes towards scalable, yet accurate methods for robustness estimation of large complex networks. 


\section{Introduction}
Man-made networks exhibit highly similar statistical characteristics, while displaying substantial non-trivial topological features~\cite{Albert2002,Barrat2004}. 
During the last decades, empirical studies have shed light on the topology of air transportation networks~\cite{Sun2014416}, electrical power grids~\cite{albert2004structural}, the Internet backbone~\cite{yook2002modeling}, inter-bank networks~\cite{boss2004network}, etc. Most of these networks carry a well-recognized resistance towards random failures~\cite{amaral2000classes} but are particularly susceptible to intentional attacks, which target relatively important nodes in the network, e.g. specific nodes that are highly connected, so-called hubs~\cite{Albert2000}. If such attacks occur, networks disintegrate rapidly. Famous examples of recent extensive, wide-ranging network failures include European air traffic disruption caused by Icelandic volcano Eyjafallaj\"okull~\cite{brooker2010fear}, large-scale power outages in the United States~\cite{andersson2005causes,ash2007optimizing}, 
Internet-based computer attack spreading~\cite{strogatz2001exploring}, and cross-continental supply-chain shortages in the Japanese tsunami aftermath~\cite{kim2015supply}. Such disruptions cause high economic costs~\cite{en81012187} and, thus, analyzing and improving network resilience remains one of the critical challenges.

In order to measure the robustness of a network, a notion of network robustness was proposed in~\cite{schneider2011mitigation}. Given a network with $N$ nodes, the robustness is defined as $R=\frac{1}{N}\sum_{Q=1}^{N}s(Q)$, where $s(Q)$ is the size of the giant component (GC size) after removing $Q$ nodes. The value of $R$ depends significantly on the underlying attacking strategy, i.e. the order in which node removals occur. Studies on network robustness usually select a single network metric, e.g. degree, betweenness, or collective influence~\cite{morone2015influence}, to generate a ranking for inducing the order. This approach has several limitations. 

\begin{figure}
\subfloat{
\includegraphics[width=.46\textwidth]{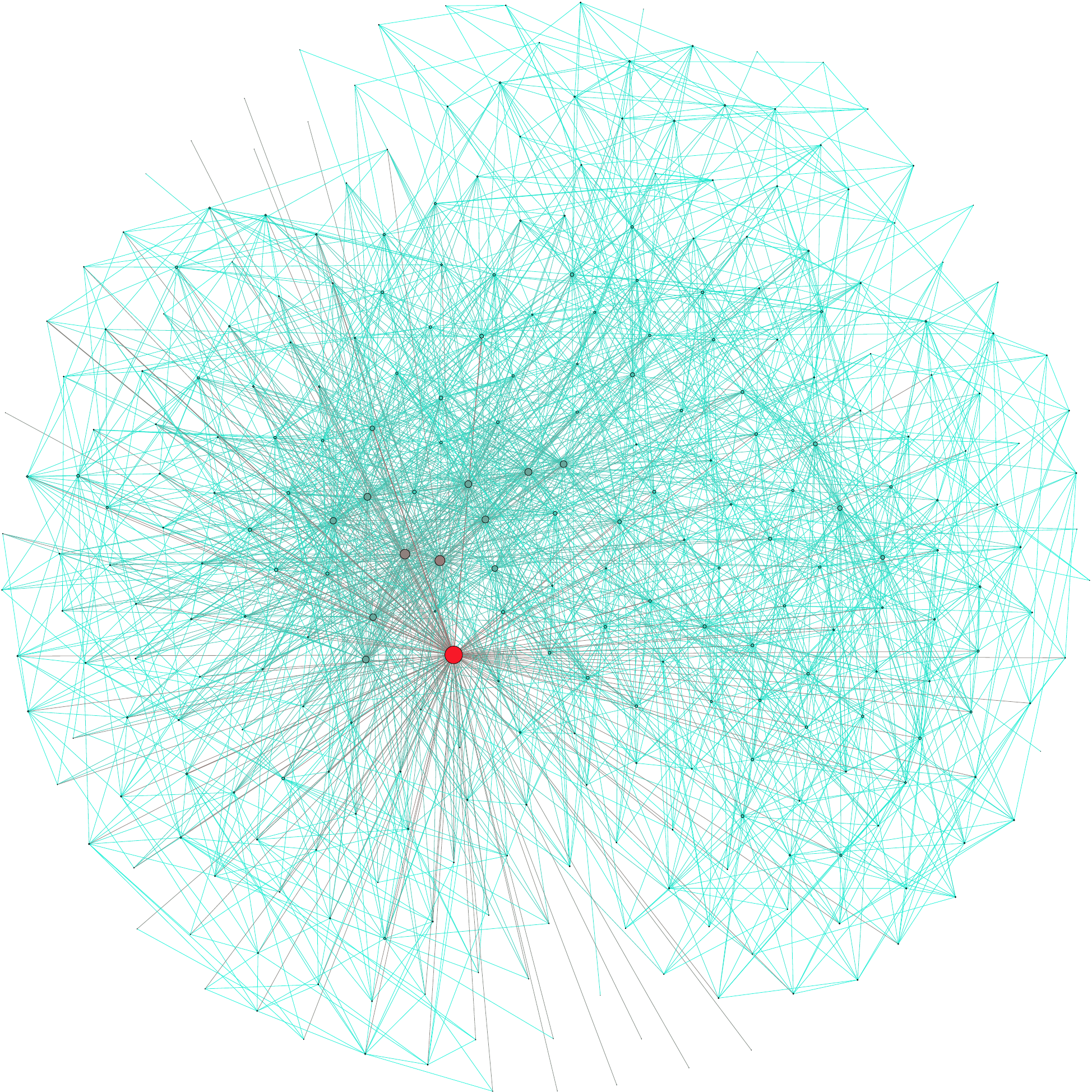}}\hfill
\subfloat{\includegraphics[width=.46\textwidth]{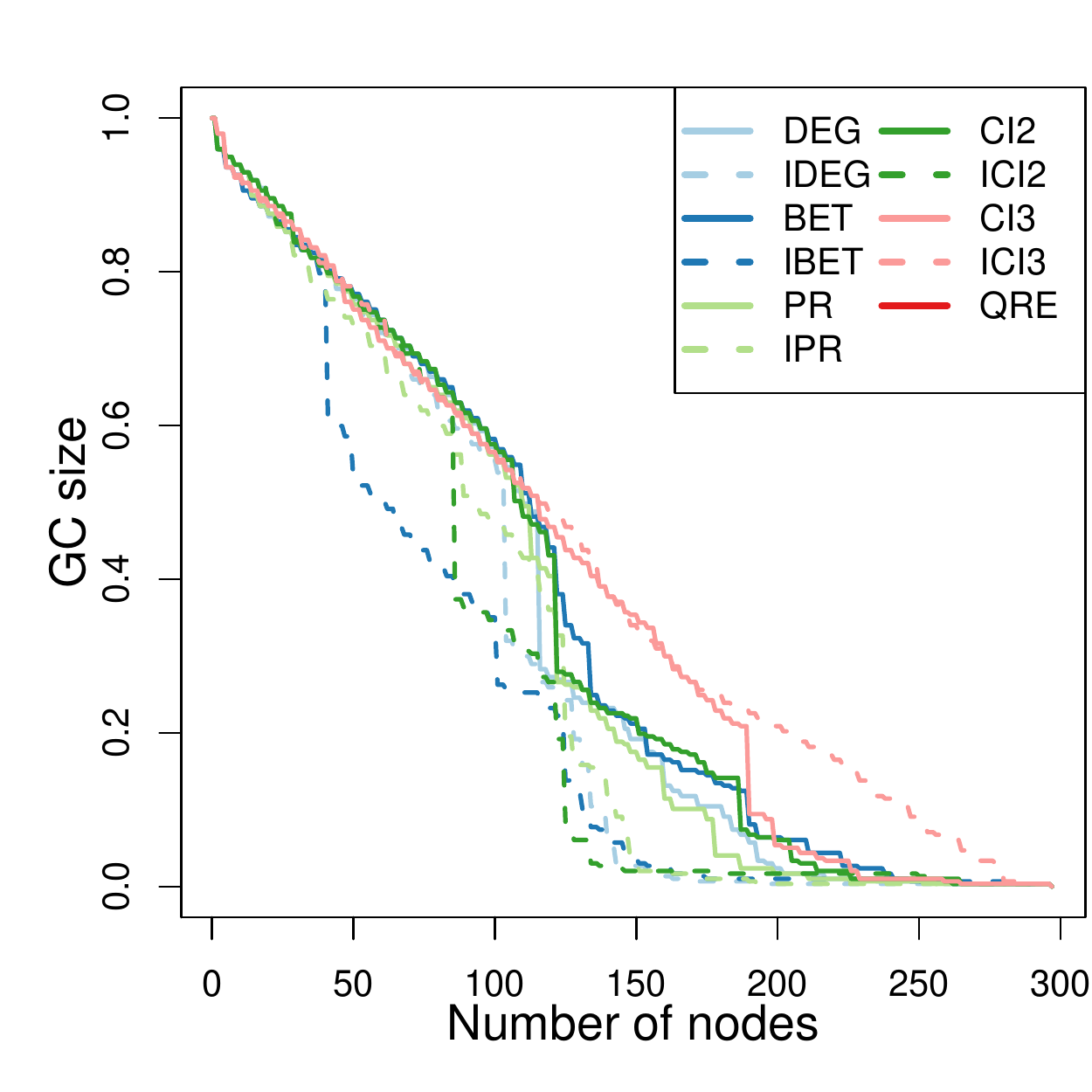}}
\caption{Visualization (left) and robustness curves induced by different network metrics (right) for the network of celegans neural. The variety of robustness curves suggests a range of R values between 0.24 (interactive betweenness, IBET) and 0.41 (Collective influence with ball size 3, ICI3). The goal is to identify a minimum R value, which is a worst case estimation of the network robustness.}
\label{fig:CelegansneuralCover}
\end{figure}

1) The obtained $R$ value reflects the robustness against a particular attacking strategy, for instance, by attacking high-degree nodes first. In Figure \ref{fig:CelegansneuralCover}, we show a visualization of the celegansneural network together with the evolution of the network's robustness (measured by GC size) while being attacked according to different rankings. We can observe a wide range of results; yielding significantly different robustness estimations of the network. In general, one is interested in the worst-case robustness, represented by the minimum R value. Note that it can be even more effective to eliminate a combination of hubs and central, but less-well-connected, nodes~\cite{kovacs2015network,nie2015new} or other nodes identified by artificial intelligent techniques~\cite{doi:10.1080/23249935.2015.1089953}. 

2) Different network metrics have significantly different computational requirements. For instance, computing the betweenness of all nodes in the network, a strategy which has been shown rather effective for attacking complex networks, needs at least time quadratic in the number of nodes. This makes it very difficult to obtain good attacks (leading to smaller R values) for very large networks with millions of nodes. For instance, the computation of betweenness values in a network with 100,000 nodes takes almost one day on today's consumer computers; and doubling the size of the network further quadruples the execution time. 

\begin{figure}[!t]
\begin{center}
\begin{tikzpicture}[scale=0.6, auto,swap]
    \foreach \pos/\name in {{(0,1)/a}, {(0,-1)/b}, {(2,0)/c},{(4,0)/d}, {(6,0)/e}, {(8,0)/f}, {(10,-1)/g}, {(10,1)/h}, {(12,0)/i}}
        \node[vertex] (\name) at \pos {$\name$};

    \foreach \source/ \dest /\weight in {a/c/, b/c/,c/d/,d/e/,e/f/,f/g/,f/h/,g/i/}
        \path[edge] (\source) -- node[weight] {$\weight$} (\dest);
        
         \node[vertex, fill=green] at (6,0) {$e$}; 
         \node[vertex, fill=cyan] at (2,0) {$c$}; 
         \node[vertex, fill=cyan] at (8,0) {$f$}; 
\end{tikzpicture}
\end{center}
\caption{Node $e$ (green) has the highest betweenness and attacking $e$ splits the network into two components with GC size four. However, when attacking two nodes, $e$ is not part of the best attack anymore, since any attack containing $e$ will necessarily have a GC size of four (only one components can be broken down further). The best attack with two nodes contains $c$ and $f$, leading to a GC size of two. Relying on rankings misses such cases.}
\label{fig:EGfig1}
\end{figure}
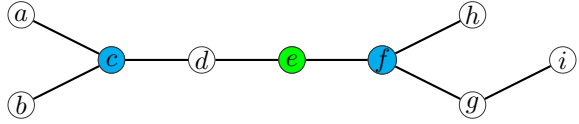

3) A fixed ranking assumes that an attack of length $n$ is a good prefix for an attack of length $n+1$. This is often not the case, as our simple example in Figure~\ref{fig:EGfig1} shows. 

4) Last but not least, existing studies often employ a fixed-length interval sampling strategy for computing the size of the giant component less than $N$ times; in order to avoid further increases on time complexities. This static sampling yields to overestimated network robustness, particularly for large vulnerable networks.

In this paper, we propose a new technique (QRE=Quick Robustness Estimation) for estimating the robustness of a network, as expressed by $R$, in subquadratic time. We assign the importance of nodes by exploiting approximate betweenness centrality, which estimates the centrality of nodes by sampling a sub-collection of node pairs. Furthermore, we iteratively adapt sampling intervals fitting the shape of the robustness curves, yielding an increasingly improved solution after each iteration. 
Experiments on real-world networks and random networks show that QRE estimates R-values close to or even better than interactive betweenness centrality, while having attractive computational properties. Our work contributes towards scalable, yet accurate methods for robustness estimation of complex networks.

\section{Methods}
\label{sec:method}

We propose QRE, an algorithm for estimating the robustness of a network, which needs less than quadratic time in the number of nodes. Conceptually, our algorithm consists of two stages:

\subsection{Initial sample shaping by degree} We obtain a first approximation of the network robustness by splitting the interval $[0,N]$ into X equi-length sub-intervals. For each such sub-interval, we attack nodes according to the decreasing degree. The metric is recomputed for each interval.
From these results, we obtain the GC size for 100 sampled points in $[0,N]$, at $p_1$ to $p_{X}$, with $gcs_1$ to $gcs_X$. 
We derive a step-function from these samples as $sf(Q)$, where $sf(Q)=gcs_i$, such that $i$ is maximal with $p_{i}\le Q$ and $p_{i+1}>Q$. We materialize this curve as a list of elements $gcs_{min}=[sf(x)\mid x\in [0,N]]$ and compute the preliminary $R$ of the network as $R_0=\frac{1}{N}\sum_{Q=1}^{N}gcs_{min}$. 

\subsection{Improvement by approximate betweenness} We split the interval $[0,N]$ into 100 equi-depth sub-intervals, according to the curve obtained by $a(Q)$. This often leads to significantly different sample points than equal-length intervals obtained in Step 1).
For each such sub-interval, we iteratively compute an attack according to the decreasing approximate betweenness of a node. In order to approximate the betweenness, we perform betweenness sampling with Y node pairs~\cite{geisberger2008better}.
From these results, we obtain the GC size for 100 sampled points in $[0,N]$, at $p_1$ to $p_{X}$, with $gcs_1$ to $gcs_X$. We combine these new samples with the previously materialized curve (representing the best known attacks so far) as follows: $gcs_{min}^{new}=[min(gcs_{min}[x],sf(x))\mid x\in [0,N]]$, where $sf(Q)=gcs_i$, such that $i$ is maximal with $p_{i}\le Q$ and $p_{i+1}>Q$. The $R$ of the network is updated as $R_0=\frac{1}{N}\sum_{Q=1}^{N}gcs_{min}$. Step 2) is repeated for $Z$ times.

\subsection{Time Complexity} This first step is performed in $\mathcal{O}(X*N*log N)$ steps, while the second step is performed in $\mathcal{O}(X*Y*Z*log N)$. For our experiments, we chose X=100, which guarantees a precision of two digits for the obtained R values. Furthermore, we let $Y=Z=sqrt(log N)$. In this way, we obtain an overall sub-quadratic time complexity of $\mathcal{O}(N*log N*log N)$.

\begin{figure*}[!t]
\begin{center}
\centerline{\includegraphics[width=1\textwidth]{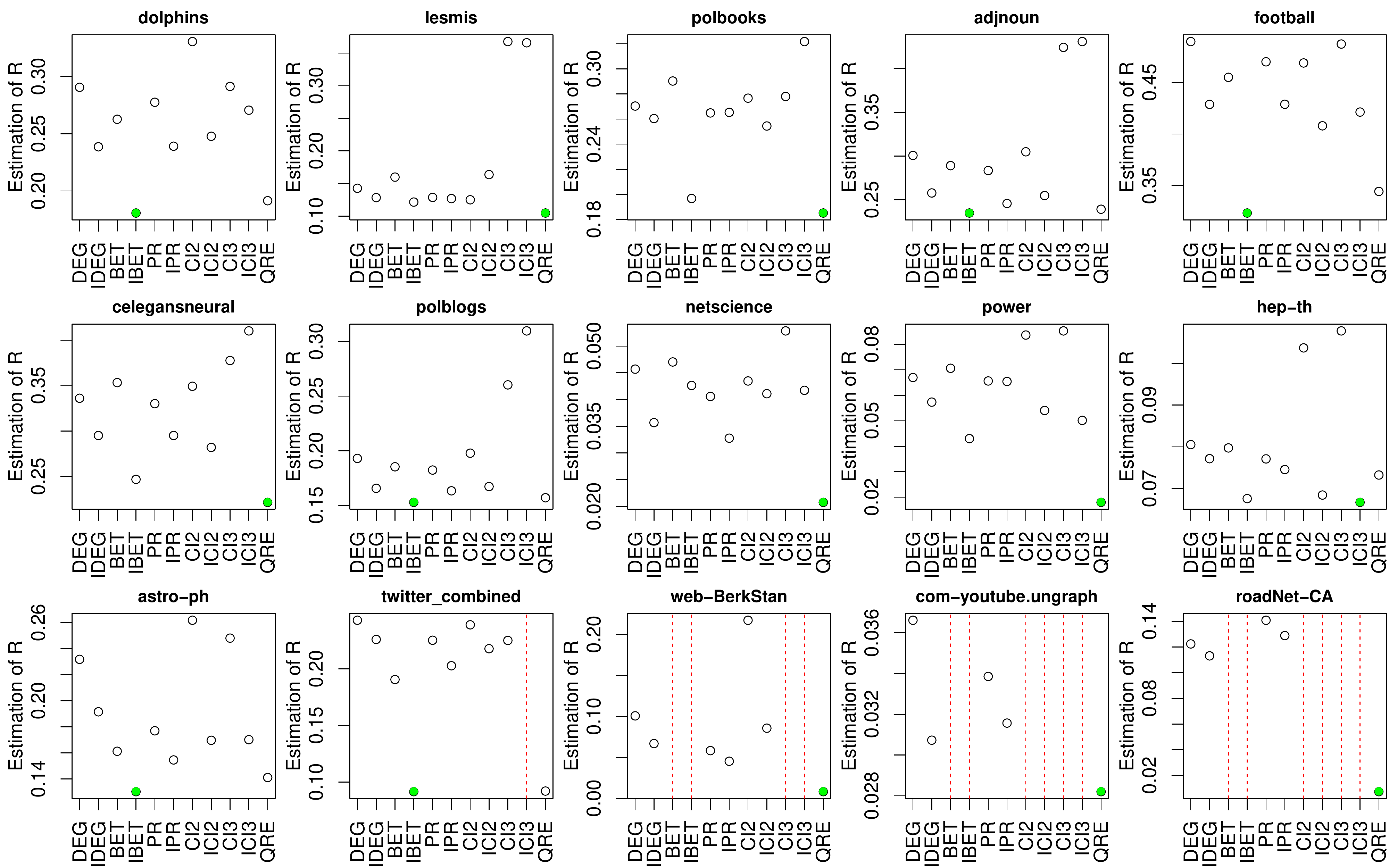}}
\caption{Robustness scores (R values) for 15 real-world networks (from UCI and SNAP) using 11 techniques.  The best technique, in terms of minimum R value, is indicated with green color. Techniques whose computations did not finish within 24 hours are marked with a red dashed line. It can be seen that QRE computes the minimum R value for the majority of cases (8/15), closely followed by interactive betweenness (6/8). For dataset hep-th, collective influence with ball size 3 computes the minimum R value.}
\label{fig:REst}
\end{center}
\end{figure*}

\section{Results}
\label{sec:Evaluation}

\subsection{Evaluation Setup and Data Sources}
We report the results of our evaluation on 15 real-world networks and two types of random networks. All experiments were executed on a server with with 32 cores and 320 GB RAM, running Fedora 21 (Linux 4.1.13-100.fc21.x86\_64). Each competitor was implemented in a single-threaded fashion, using the network library NetworKit\footnote{Available at https://networkit.iti.kit.edu/}. All datasets are available to download from either the UCI Network Data Repository \footnote{Available at https://networkdata.ics.uci.edu/index.php} 
or the Stanford Large Network Dataset Collection 
\footnote{Available at https://snap.stanford.edu/data/}, and have been studied extensively. They cover a variety of network structures and network scales, ranging from a dozen of nodes (dolphins, lesmis) to more than a million of nodes (com-youtube.ungraph, roadNet-CA). Moreover, we have performed experiments with two types of random networks (Barabási-Albert and Erd\"os-Renyi).

\begin{table}[]
\centering
\caption{List of 15 datasets used in our experiments, ranked with an increasing number of nodes. Robustness estimations are based on interactive betweenness; only results obtained within 24 hours of computation are shown (others are marked with $*$).}
\label{my-label}
\begin{tabular}{|l|r|r|r|r|}
\hline
Network             & |N|     & |L|     & Est. R   & Time in s \\ \hline
dolphins            & 62      & 159     & 0.18      & 0.07                \\
lesmis              & 77      & 254     & 0.12      & 0.09                \\
polbooks            & 105     & 441     & 0.20      & 0.17                \\
adjnoun             & 112     & 425     & 0.23      & 0.27                \\
football            & 115     & 613     & 0.32      & 0.23                \\
celegansneural      & 297     & 2148    & 0.25      & 0.79                \\
polblogs            & 1490    & 16715   & 0.15      & 10.59               \\
netscience          & 1589    & 2742    & 0.04      & 0.93                \\
power               & 4941    & 6594    & 0.04      & 25.50               \\
hep-th              & 8361    & 15751   & 0.07       & 67.34               \\
astro-ph            & 16706   & 121251  & 0.13       & 1076.38             \\
twitter\_combined   & 81306   & 2420765 & 0.09      & 65666.80            \\
web-BerkStan        & 685230  & 7600595 & $*$      & $*$                  \\
com-youtube.ungraph & 1134890 & 2987624 & $*$       & $*$                  \\
roadNet-CA          & 1965206 & 5533214 & $*$      & $*$  \\
\hline
\end{tabular}
\end{table}

\subsection{Robustness estimations for real-world networks}
\begin{figure}[!t]
\begin{center}
\centerline{\includegraphics[width=0.90\textwidth]{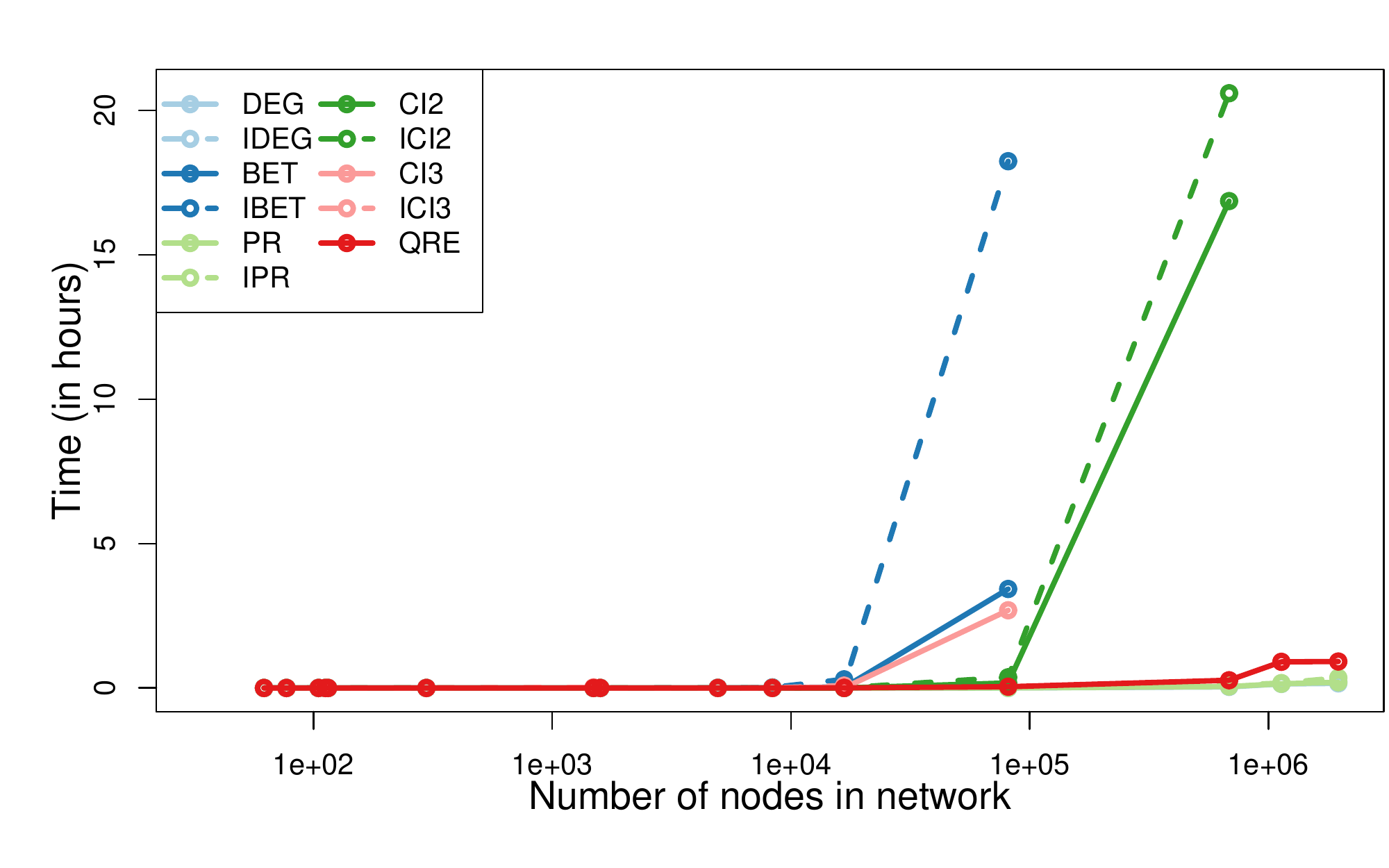}}
\caption{Execution time for all techniques against real-world networks, with a growing number of nodes. IBETW and BETW scale worst, followed by instances of collective influence. Only QRE and the simple methods based on degree and pagerank can compute the R values for networks with more than a million of nodes within reasonable time (less than one hour).}
\label{fig:Time}
\end{center}
\end{figure}

\begin{figure}[!t]
\begin{center}
\centerline{\includegraphics[width=0.80\textwidth]{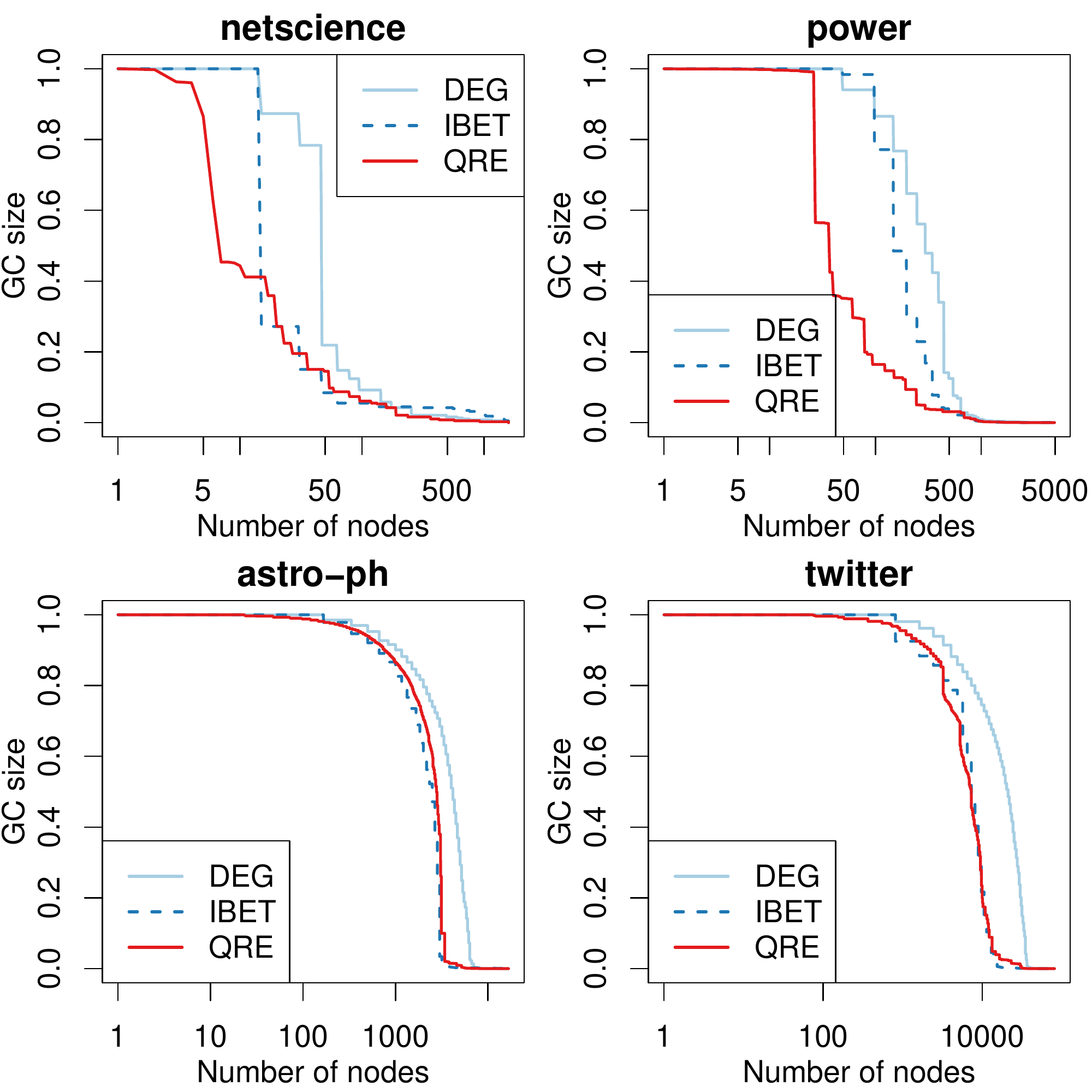}}
\caption{Size of the giant component for varying number of nodes according to static degree (DEG), interactive betweenness (IBET) and QRE. Attacks identified by QRE break down the network significantly earlier than IBET, for networks netscience and power (upper row). For large networks (astro-ph and twitter\_combined, lower row), IBET and QRE are competitive with each other.}
\label{fig:PowerComparison}
\end{center}
\end{figure}

\begin{figure}[!t]
\begin{center}
\centerline{\includegraphics[width=.90\textwidth]{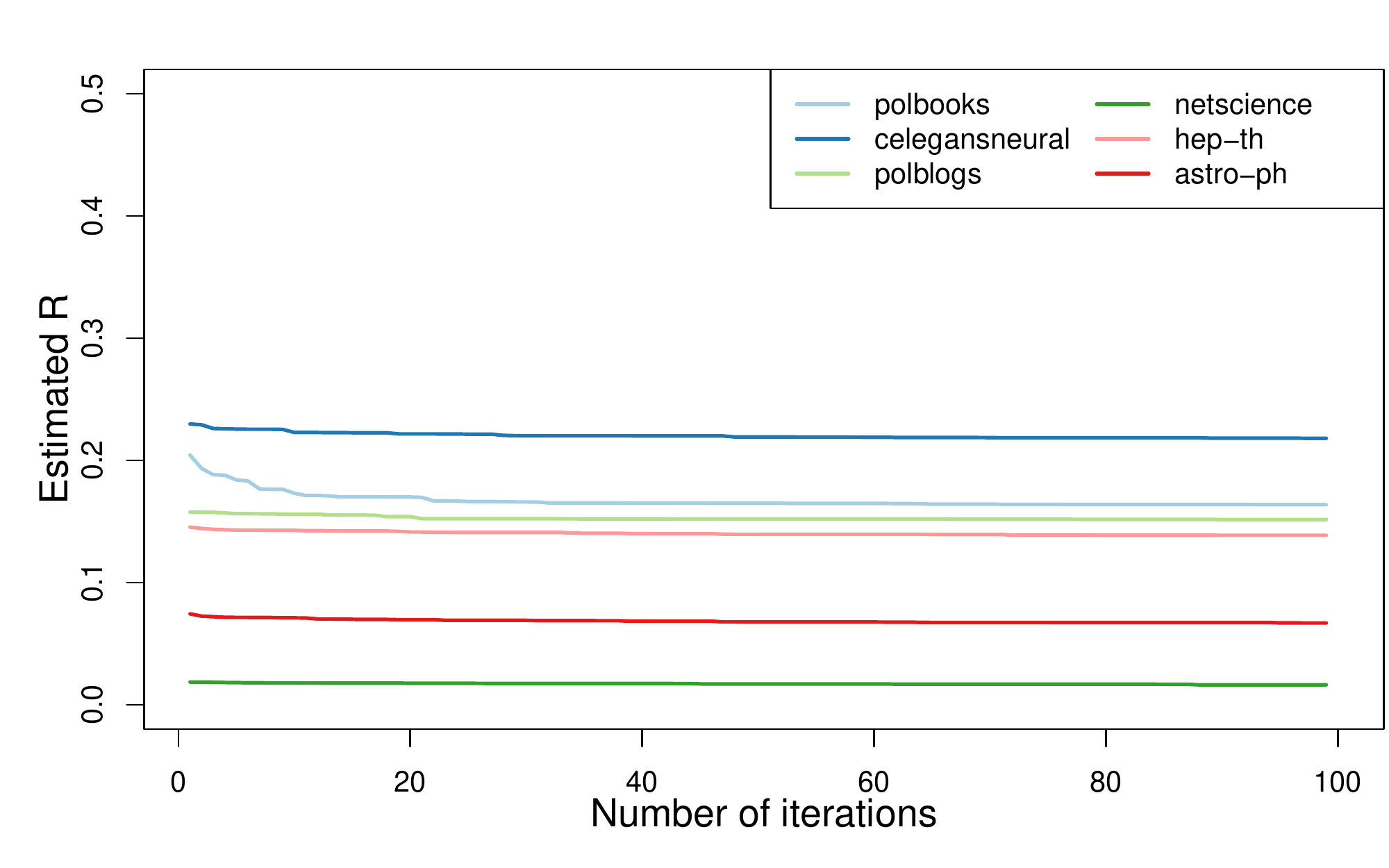}}
\caption{Estimated R value for 1--100 iterations of QRE and six real-world datasets. For most networks, the estimated robustness is stable after a few initial iterations; only network polbooks needs approx. 10 iterations to reach a fixed point. Iterating more times does not change the results significantly.}
\label{fig:IterationsVisualization}
\end{center}
\end{figure}

In Figure~\ref{fig:REst}, we present the estimated robustness (R values) for all 15 networks computed by different techniques: DEG (=degree), BETW (betweenness), PR (pagerank), CI2 (collective influence~\cite{morone2015influence} with ball size 2), CI3 (collective influence with ball size 3), and QRE (the approach proposed in this study). The symbol 'I' in front of a technique's name indicates an interactive attack, where the node ranking is recomputed after attacking a prefix of nodes. It can be seen that QRE computes the minimum R value for the majority of cases (8/15), followed by interactive betweenness (6/15). In Figure~\ref{fig:Time}, we report the running times using the 11 techniques on real-world networks. Smaller networks with less than few hundred nodes can be analyzed by all techniques within a few seconds. However, as the size of the network grows, several of them do not scale well anymore. IBET needs already 18 hours for twitter\_combined with approx. 80000 nodes. Collective influence with ball size 2 needs 16-20 hours for dataset web-BerkStan; Collective influence with ball size 3 scales even worse, since more neighbors need to be traversed. The simple techniques based on degree and pagerank scale well with the number of nodes, outperforming QRE slightly. 
In Figure~\ref{fig:PowerComparison}, we compare the size of the giant component against a number of attacked nodes for the three most interesting techniques: Static degree (fast to compute), Interactive betwenneess (small R values, but time consuming), and QRE (trade-off). In Figure~\ref{fig:IterationsVisualization}, we evaluate the influence of the number of iterations on the $R$ value of a network. Usually, after a few number of iterations, the value is not decreased further significantly.

\subsection{Robustness estimations for random networks} 
\begin{figure}[!t]
\begin{center}
\centerline{\includegraphics[width=.90\textwidth]{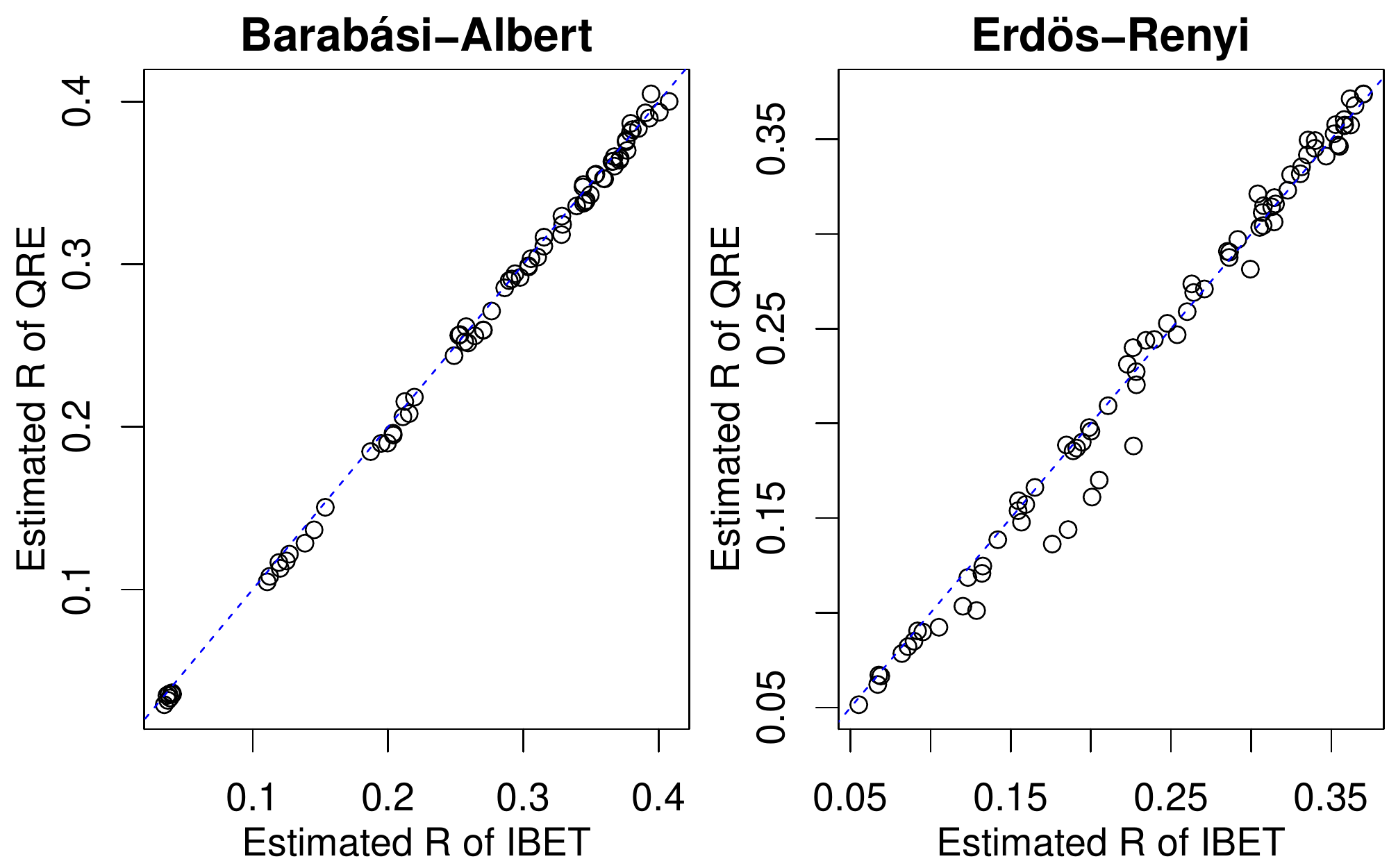}}
\caption{Scatter plot of R values obtained by QRE against IBET for 100 artificial networks with 500 nodes, generated by Barabási-Albert model (left) and Erd\"os-Renyi model (right). A linear correlation between R values can be observed, suggesting that QRE also accurately estimates the robustness of artificial networks.}
\label{fig:ANM}
\end{center}
\end{figure}

We have performed additional experiment with two random graph models: Barabási-Albert and Erdös-Renyi. In Figure~\ref{fig:ANM}, we report the estimated robustness of 100 artificial networks, for interactive betweenness (IBET) and our novel technique (QRE). There exists a strong correlation between both scores, indicating that QRE accurately estimates the robustness even for random networks. In few cases, for networks with $0.15\le R\le 0.25$, IBET finds even better attacks than QRE.

\section{Discussion}

Our robustness estimation technique, QRE, scales up to large networks with millions of nodes and edges, since the overall time complexity is in $\mathcal{O}(N*log N*log N)$, where $N$ is the number of nodes in the network. Based on our experiments with real-world and artificial networks, we conclude that QRE fills a nice sweet spot between fast, yet ineffective degree centrality and slow, yet highly effective betweenness centrality. The results of our study show that efficient, yet accurate robustness estimation is possible even for very large networks. We believe that this work contributes to a better understanding of real-world network robustness in face of big data. Moreover, we envision that our technique can be extended to analyze robustness of networks of networks~\cite{gao2014single}.

\section{Acknowledgment}
The authors would like to acknowledge very fruitful discussions with Shlomo Havlin (Bar-Ilan University) and Massimiliano Zanin (Innaxis Foundation \& Research Institute) which significantly improved this manuscript. 

\bibliography{document}
\bibliographystyle{abbrv}
\end{document}